\documentstyle[preprint,aps,epsfig]{revtex}

\def\beq{\begin{equation}}
\def\eeq{\end{equation}}
\newcommand{\bscco}{{Bi$_2$Sr$_2$Ca$_2$Cu$_2$O$_{8+\delta}$}}

\tightenlines
\begin{document}

\title{Models for Enhanced Absorption in Inhomogeneous Superconductors}
\author{Sergey\ V.\ Barabash, David Stroud}
\address{Department of Physics,
The Ohio State University, Columbus, Ohio 43210}
\date{\today}

\date{\today}

\maketitle

\begin{abstract}

We discuss the low-frequency absorption arising from quenched 
inhomogeneity in the superfluid density $\rho_s$ of a 
model superconductor.  Such inhomogeneities 
may arise in a high-T$_c$ superconductor from a wide variety of
sources, including quenched random disorder and static 
charge density waves
such as stripes.  Using standard classical methods for treating
randomly inhomogeneous media, we show that both 
mechanisms produce additional absorption at finite 
frequencies.  For a two-fluid
model with weak mean-square fluctuations 
$\langle(\delta\rho_s)^2\rangle$ in $\rho_s$
and a frequency-independent quasiparticle conductivity,
the extra absorption has oscillator strength proportional to
the quantity
$\langle(\delta\rho_s)^2\rangle/\rho_s$, as observed in some
experiments.  Similar behavior is found in a two-fluid model with
anticorrelated fluctuations in the superfluid and normal fluid
densities.  The extra absorption typically occurs as a Lorentzian
centered at zero frequency.  We present simple model calculations
for this extra absorption under conditions of 
both weak and strong fluctuations.  The relation between our
results and other model calculations is briefly discussed.

\end{abstract}

\draft \pacs{PACS numbers: }
 
\par

\newpage

\section{Introduction}

Some high-T$_c$ superconductors, such as \bscco,  
absorb very strongly in the microwave regime even at temperatures
$T$ far below $T_c$\cite{corson,lee}.  At a fixed frequency as
a function of $T$, the real conductivity typically has two 
features: a sharp peak near $T_c$, and a broad, 
frequency-dependent background extending to quite low $T$.
The peak near T$_c$ is usually
ascribed to critical fluctuations arising from the 
superconducting transition.

The origin of the broad background is less clear.  In 
conventional s-wave superconductors, there is no such
background, because the material cannot 
absorb radiation at frequencies below
the energy gap for pair excitations, $2\Delta$.  
But in high-T$_c$ materials, which are thought to 
have a $d_{x^2-y^2}$ order parameter\cite{vanharlingen,kirtley}, 
the gap vanishes in certain nodal directions in $k$-space.  
Hence, gapless nodal quasiparticles can be excited at arbitrarily
low $T$, and these can absorb at low frequencies.
But the observed absorption appears to be stronger 
than expected from the quasiparticles alone in a 
two-fluid model\cite{corson2}.  

Several authors have suggested quenched
inhomogeneities as the origin of this extra absorption.  
Such inhomogeneities could arise from statistical fluctuations
in the local densities of holes or impurities, or from the
presence of superfluid-suppressing impurities such as Zn.   
Corson {\it et al}\cite{corson} found that such 
fluctuations in \bscco 
displace about 30\% 
of the spectral weight from the
condensate to the conductivity at finite frequencies.  
In addition, several explicit models have been
presented which produce such a displacement.  For example,
Van der Marel and Tsvetkov\cite{marel} have calculated
the fluctuation conductivity due to a periodic one-dimensional
modulation of the phase stiffness.
The present authors\cite{barabash} have calculated 
the fluctuation conductivity of a two-dimensional array
of Josephson junctions with quenched randomness in the critical
currents, using the Kubo formalism\cite{kubo}.  They found
an excess contribution to the conductivity below $T_c$ arising
from this inhomogeneity, which arose from displacement of some
of the superfluid contribution from zero to finite frequencies.

More recently, Orenstein\cite{orenstein} considered a 
two-fluid model of superconductivity with quenched, anticorrelated 
inhomogeneity in both the normal and superfluid densities;
he found that some of the spectral weight of the conductivity
is displaced from zero to finite frequencies, and gave explicit 
expressions for this extra conductivity.
Han\cite{han} has given a field-theoretic treatment of a
randomly inhomogeneous two-dimensional d-wave superconductor,
using a replica formalism to treat the assumed weak disorder.
He obtained an extra Lorentzian peak in the real conductivity, 
centered at zero frequency, arising from this disorder,
and a corresponding reduction in the superfluid density.  All
such models are made more plausible by imaging
experiments which have directly observed spatial 
fluctuations in the local superconducting
energy gap in underdoped cuprate superconductors\cite{lang}.

In this paper, we study the low-frequency complex conductivity
of several models for two-dimensional superconductors with 
quenched inhomogeneities.  We use a straightforward approach
to obtain the effective complex conductivity, based on standard
treatments of randomly inhomogeneous media in the 
quasistatic regime\cite{bergman}.  From this approach, we obtain
sum rules for the complex conductivity.  We also find, in a number
of cases, explicit expressions for the extra conductivity
$\delta\sigma_e(\omega)$ due to the quenched inhomogeneities.  
Where the models are comparable, our 
explicit form for $\delta\sigma_e(\omega)$ reduces to that
of Han, even though obtained using a quite different approach.

The remainder of this paper is organized as follows.  In Section
II, we develop sum rules for the optical conductivity
of a two-dimensional superconductor with quenched inhomogeneities,
and apply them to several explicit models.
In Section III, we present some approximate
calculations for the effective conductivity of such 
superconductors\cite{note:etopim}.
A concluding discussion is given in Section IV.

\section{Conductivity Sum Rules} 

\subsection{Kramers-Kronig Relations}

We consider a two-dimensional superconductor, in which
the local conductivity has normal and superfluid 
contributions which act in parallel, but,
in contrast to the usual two-fluid model, these terms
are spatially varying.  The complex local conductivity
$\sigma({\bf x}, \omega)$ is taken as a scalar function
of position ${\bf x}$ and frequency $\omega$ of the following form:
\begin{equation}
\sigma(\omega, {\bf x}) =  \sigma_n(\omega, {\bf x}) + 
\frac{i\rho_s({\bf x})}{\omega}.
\label{eq:model}
\end{equation}
$\sigma_n$ might be a contribution from the nodal 
quasiparticles expected in a superconductor with $d_{x^2-y^2}$ order 
parameter symmetry, while $\rho_s$ represents the perfect-conductivity response of the
superconductor.  
Our goal is to calculate the position-independent
effective complex conductivity of
this medium, which we denote
\begin{equation}
\sigma_e(\omega) \equiv \sigma_{e,1}(\omega) + i\sigma_{e,2}(\omega).
\end{equation}


We first obtain a sum rule for $\sigma_{e,1}$.  
We start from the Kramers-Kronig relations obeyed by any function,
such as $\sigma_e(\omega)$, which describes a causal response:
\begin{equation}
\sigma_{e,1}(\omega) = \frac{2}{\pi}P\int_0^\infty
\frac{\omega^\prime \sigma_{e,2}(\omega^\prime)}
{\omega^{\prime 2} - \omega^2}d\omega^\prime + \sigma_\infty
\label{eq:kk1}
\end{equation}
and
\begin{equation}
\sigma_{e,2} = \frac{\rho_{s,e}}{\omega} - \frac{2\omega}{\pi}
P\int_0^\infty\frac{\sigma_{e,1}(\omega^\prime)-\sigma_\infty}
{\omega^{\prime 2} - \omega^2}d\omega^\prime.
\label{eq:kk2}
\end{equation}
Here P denotes the principal part, and
$\sigma_\infty \equiv$ 
Lim$_{\omega\rightarrow\infty}\sigma_{e,1}(\omega)$.  We have assumed that
Lim$_{\omega\rightarrow\infty}\sigma_n(\omega)$ 
position-independent, and have used the fact that,
at high frequencies, $\sigma_{e,2} \rightarrow i\rho_{s,e}/\omega$,
where $\rho_{s,e}$ is the effective superfluid density.
At sufficiently large $\omega$, the right hand side
of eq.\ (\ref{eq:kk2}) takes the form
\begin{equation}
\sigma_{e,2} \rightarrow \frac{1}{\omega}\left(\rho_{s,e} + 
\frac{2}{\pi}\int_0^\infty
[\sigma_{e,1}(\omega^\prime) - \sigma_\infty]d\omega^\prime\right).
\label{eq:kk3}
\end{equation}  

We first write down a perturbation result for
$\sigma_e(\omega)$, which will be needed below.  If the spatial
fluctuations in $\sigma(\omega, {\bf x})$ are small in magnitude
compared to it spatial average, $\sigma_{av}$, 
then to second order in the 
$|\delta\sigma|/|\sigma_{av}| \equiv 
|\sigma(\omega, {\bf x}) -
\sigma_{av}(\omega)|/|\sigma_{av}(\omega)|$,
\begin{equation}
\sigma_e \approx \sigma_{av} -\frac{1}{2}
\frac{\langle (\delta\sigma)^2\rangle}{\sigma_{av}},
\label{eq:weak}
\end{equation}
where $\langle ...\rangle$ denotes a space average
of the quantity in brackets\cite{bergman}.

\subsection{Frequency-Independent Normal Conductivity}

Suppose first that $\sigma_n(\omega)$ is real, nonzero,
frequency-independent, and spatially uniform.
Then $\sigma_\infty = \sigma_n$, and, at
high $\omega$, however large the fluctuations
in $\rho_s({\bf x})$, the fluctuations in the 
{\em local} complex conductivity,
$\sigma_n + i\rho_s({\bf x})/\omega$, 
are small compared to the space-averaged conductivity
$\sigma_{av}(\omega) = \sigma_n + i\rho_{s,av}/\omega$; we can
thus use eq.\ (\ref{eq:weak}). 
Since only $\rho_s$ is fluctuating, and not $\sigma_n$, 
eq.\ (\ref{eq:weak}) becomes
\begin{equation}
\sigma_e \approx \sigma_n + \frac{i\rho_{s,av}}{\omega} + 
\frac{1}{2}\left(\frac{\langle(\delta \rho_s)^2\rangle}
{\omega^2\sigma_n + i\omega \rho_{s,av}}\right),
\label{eq:hifreq}
\end{equation}
where $\delta\rho_s({\bf x}) = \rho_s({\bf x}) - \rho_{av}$.
Thus, to leading order in $1/\omega$,
$\sigma_{e,2} \sim \rho_{s,av}/\omega$.
Equating this expression to the right-hand
side of eq.\ (\ref{eq:kk3}), we obtain
\begin{equation}
\int_0^\infty
\left[\sigma_{e,1}(\omega^\prime)-\sigma_n\right]d\omega^\prime
= \frac{\pi}{2}\left(\rho_{s,av} - \rho_{s,e}\right)
\label{eq:sige1}
\end{equation}
One consequence of eq.\ (\ref{eq:sige1}) is that, 
if $\rho_s({\bf x})$ is spatially varying, 
$\rho_s({\bf x})$, $\rho_{s,av} > \rho_{s,e}$, 
the right-hand side of eq.\ (\ref{eq:sige1}) is positive,
and {\em there will be an additional contribution to 
$\sigma_{e,1}(\omega)$, above $\sigma_n$}.  

The sum rule (\ref{eq:sige1}) requires
only that the conductivity fluctuations
are asymptotically small at large $\omega$, and 
not necessarily that $\rho_s$ have small fluctuations.
If, however, the fluctuations in $\rho_s$ are, in fact, small,
then $\rho_{s,e}$ can be calculated approximately using the
analog of eq.\ (\ref{eq:weak}), namely,
\begin{equation}
\rho_{s,e} \approx \rho_{s,av} - 
\frac{1}{2}\frac{\langle(\delta\rho_s)^2\rangle}{\rho_{s,av}}.
\label{eq:rhose}
\end{equation}
From this approximate result, eq.\ (\ref{eq:sige1}) becomes
\begin{equation}
\int_0^\infty\left[\sigma_{e,1}(\omega^\prime)-\sigma_n\right]
d\omega^\prime
\sim \frac{\pi}{4}\rho_{s,av}\left(\frac{\langle(\delta \rho_s)^2\rangle}
{\rho_{s,av}^2}\right).
\label{eq:sige1p}
\end{equation}
For fixed mean-square fluctuations 
$\langle (\delta \rho_s)^2\rangle/\rho_{s,av}^2$,
this integral is proportional to the average superfluid density $\rho_{s,av}$.
That is, the extra integrated fluctuation contribution to 
$\sigma_{e,1}$, 
above the mean contribution of the normal fluid, 
is proportional to $\rho_{s,av}$.  
A similar result has been reported in experiments 
carried out over a range of mean superfluid 
densities\cite{corson}.

\subsection{Frequency-Dependent, Spatially Fluctuating Normal
Conductivity}

Next, we consider a frequency-dependent
$\sigma_n(\omega, {\bf x})$, assuming that
Lim$_{\omega\rightarrow\infty}\sigma_n(\omega,{\bf x}) = 0$, 
but allowing for a spatially varying 
$\sigma_n(\omega, {\bf x})$.
Then eq.\ (\ref{eq:kk2}) becomes
\begin{equation}
\sigma_{e,2}(\omega) = \frac{\rho_{s,e}}{\omega} - 
\frac{2\omega}{\pi}
P\int_0^\infty\frac{\sigma_{e,1}(\omega^\prime)}
{\omega^{\prime 2}-\omega^2}d\omega^\prime
\rightarrow \frac{1}{\omega}\left(\rho_{s,e}+\frac{2}{\pi}
\int_0^\infty\sigma_{e,1}(\omega^\prime)d\omega^\prime\right),
\label{eq:sige2pp}
\end{equation}
where the last result is again 
valid at sufficiently large $\omega$.
We now assume a Drude form for $\sigma_n(\omega, {\bf x})$, i.\ e.,
\begin{equation}
\sigma_n(\omega, {\bf x}) = 
\frac{\rho_n({\bf x})\tau_n({\bf x})}{1-i\omega\tau_n({\bf x})},
\label{eq:Drude}
\end{equation}
where $\rho_n({\bf x})$ is a suitable normal fluid density, and 
$\tau_n({\bf x})$ is
a quasiparticle relaxation time.  
Then at sufficiently high $\omega$,
\begin{equation}
\sigma_2(\omega, {\bf x}) \sim  
\frac{i[\rho_s({\bf x})+\rho_n({\bf x})]}{\omega},
\end{equation}
while $\sigma_1(\omega, {\bf x})$ falls off at least as fast
as $1/\omega^2$.
Similarly, $\sigma_{e,1}$ also
falls off at least as fast as $1/\omega^2$ at large $\omega$.  
Hence, $\sigma_e \sim i\sigma_{e,2}$ at high 
frequencies, where
\begin{equation}
\sigma_{e,2}(\omega) = \frac{(\rho_s + \rho_n)_e}{\omega}.
\label{eq:14}
\end{equation}
Here $(\rho_s + \rho_n)_e$ denotes the effective superfluid
density of a hypothetical material whose spatially varying local
superfluid density is $\rho_s({\bf x}) + \rho_n({\bf x})$.  
Equating the coefficients of $\omega^{-1}$ on 
right hand sides of eqs.\ (\ref{eq:sige2pp}) and
(\ref{eq:14}) yields
\begin{equation}
\int_0^\infty\sigma_{e,1}(\omega^\prime)d\omega^\prime =
\frac{\pi}{2}\left[(\rho_s+ \rho_n)_e  - \rho_{s,e}\right],
\end{equation}
which gives the integrated spectral weight of $\sigma_{e,1}(\omega)$.  

To calculate this weight, we first
introduce $\sigma_{n,av} \equiv \langle\sigma_n\rangle$, and
note that 
\begin{equation}
\int_0^\infty\sigma_{n,1,av}(\omega^\prime) d\omega^\prime
 = \frac{\pi\rho_{n,av}}{2},
\end{equation}
even if $\tau_n({\bf x})$ is position-dependent.
Hence, 
\begin{equation}
\int_0^\infty\left[\sigma_{e,1}(\omega^\prime) -
\sigma_{n,1,av}(\omega^\prime)\right]
d\omega^\prime = \frac{\pi}{2}\left[(\rho_s+ \rho_n)_e 
-\rho_{s,e}-\rho_{n,av}\right].
\label{eq:sigesn}
\end{equation}
Thus $\sigma_{e,1}(\omega^\prime)$ again has some extra spectral
weight beyond that of
$\sigma_{n,av}(\omega)$.  

We first estimate expression 
(\ref{eq:sigesn}) in the small-fluctuation regime where 
$|\delta \rho_s| \ll \rho_{s,av}$, assuming that
$\rho_n$ is non-fluctuating.  Following steps analogous
to those leading up to eq.\ (\ref{eq:sige1p}),
we obtain
\begin{equation}
\int_0^\infty\left[\sigma_{e,1}(\omega^\prime)-\sigma_{n,1,av}(\omega^\prime)
\right]d\omega^\prime \rightarrow 
\frac{\pi}{4}\langle(\delta \rho_s)^2\rangle
\left(\frac{1}{\rho_{s,av}}-\frac{1}{\rho_{s,av} + \rho_n}\right).
\end{equation}
In the limit $\rho_n \gg \rho_{s,av}$, 
this expression reduces to the right-hand side of 
eq.\ (\ref{eq:sige1p}). In this regime, the extra spectral weight 
is again proportional to
$\rho_{s,av}$.  A frequency-independent $\sigma_n$
is a special case of this limiting behavior.
In the opposite limit ($\rho_{s,av} \gg \rho_n$),
the behavior is
\begin{equation}
\int_0^\infty\left[\sigma_{e,1,av}(\omega^\prime)-\sigma_{n,1}(\omega^\prime)\right]
d\omega^\prime \rightarrow \frac{\pi}{4}\rho_n
\frac{\langle(\delta \rho_s)^2\rangle}{\rho_{s,av}^2}.
\label{eq:sumrule2}
\end{equation}
Thus, in this limit, for fixed mean-square fluctuations in $\rho_s$,
the extra spectral weight is {\em not} proportional to 
$\rho_{s,av}$\cite{remark}.  

If both $\rho_s$ and $\rho_n$ are spatially varying, eq.\
(\ref{eq:sigesn}) can still be easily treated
when spatial fluctuations in $\rho_s$ and $\rho_n$ are small.  
Then, to second order in these fluctuations,
\begin{equation}
(\rho_s + \rho_n)_e \approx \rho_{s,av} + \rho_{n,av}
- \frac{1}{2}\frac{\langle[\delta(\rho_s + \rho_n)]^2\rangle}
{\rho_{s,av} + \rho_{n,av}},
\label{eq:rhoe1}
\end{equation}
while $\rho_{s,e}$ is given approximately by eq.\ (\ref{eq:rhose}).
Substituting eqs.\ (\ref{eq:rhoe1}) and (\ref{eq:rhose})
into eq.\ (\ref{eq:sigesn}) gives
\begin{equation}
\int_0^\infty[\sigma_{e,1}(\omega^\prime) - \sigma_{n,1,av}
(\omega^\prime)]d\omega^\prime 
= \frac{\pi}{4}\left[\frac{\langle(\delta\rho_s)^2\rangle}{\rho_{s,av}}
- \frac{\langle[\delta(\rho_s+\rho_n)]^2\rangle}{\rho_{s,av}
+ \rho_{n,av}}\right].
\label{eq:flucn}
\end{equation}

To interpret eq.\ (\ref{eq:flucn}),
we assume that $\rho_s$ and $\rho_n$ are correlated
according to $\rho_n = \lambda\rho_s$.  Thus
$\lambda = 1$ and $\lambda = -1$ represent perfect 
correlation and perfect anticorrelation.  The predictions of
eq.\ (\ref{eq:flucn}) are particularly striking
in the latter case, as postulated in some models of
inhomogeneity\cite{orenstein,han}.  
In this case the {\em sum} $\rho_s + \rho_n$ is {\em not}
fluctuating.  Then the second term on the
right-hand side of eq.\ (\ref{eq:flucn}) vanishes, and 
the extra spectral weight due to the inhomogeneity is
again proportional to $\langle(\delta\rho_s)^2\rangle/\rho_{s,av}$.
In the opposite case $\lambda = 1$, the right-hand side of
eq.\ (\ref{eq:flucn}) may be negative, i. e. the
total spectral weight is {\em smaller}
than that given by $\sigma_{n,1,av}$.

\subsection{Tensor Conductivity}

The preceding arguments apply equally well to a {\em tensor}
conductivity, as expected if the superconducting layer 
contains charge stripes or other types 
of static charge density waves.  In this case, the
local conductivity should be a $2 \times 2$ tensor\cite{kivelson} 
of the form $\sigma^{\alpha\beta}(\omega, {\bf x}) = 
R^{-1}({\bf x})\sigma^d(\omega)R({\bf x})$,
where $\sigma^d(\omega)$ is a diagonal $2 \times 2$ matrix 
with diagonal components $\sigma_{A}(\omega)$, $\sigma_{B}(\omega)$,
and $R({\bf x})$ is a $2\times 2$ rotation matrix 
describing the local orientation of the stripes.  
If this phase has a domain
structure, the stripes would have either of 
two orientations relative to the layer crystalline axes, 
with equal probability.  In a macroscopically isotropic layer,
this domain structure leads to an scalar effective 
conductivity $\sigma_e(\omega)$. 
If, for example, the conductivity along the j$^{th}$ 
principal axis is $\sigma_{j}(\omega) = \sigma_{n,j}(\omega) +
i\rho_{s,j}/\omega$, then the superfluid inhomogeneity
produces an extra spectral weight at finite $\omega$ given by
eq.\ (\ref{eq:sigesn}).

\section{Model Calculations}

\subsection{Weak Inhomogeneity}

We now supplement
these sum rules with some
explicit expressions for $\sigma_e(\omega)$, beginning with
weak inhomogeneity, $|\delta\sigma| \ll \sigma_{av}$.  In this case, 
$\sigma_e$ is given by eq.\ (\ref{eq:weak}).  If only $\rho_s$
is spatially fluctuating, this equation reduces to eq.\
(\ref{eq:hifreq}) and 
\begin{equation}
\sigma_{e,1}(\omega) \approx \sigma_n
+ \frac{1}{2}\langle (\delta\rho_s)^2\rangle\frac{\sigma_n}
{\rho_{s,av}^2 + \omega^2\sigma_n^2}.
\label{eq:weak1}
\end{equation}
Thus, the extra absorption is a
Lorentzian centered at $\omega = 0$, of half-width
$\propto \rho_s/\sigma_n$, and strength 
$\propto\langle(\delta\rho_s)^2\rangle$.   A result equivalent to
eq.\ (\ref{eq:weak1}) has also been obtained by Han\cite{han}
using a replica formalism.  Indeed, even the parameters of
the two results are identical, provided we interpret 
$\langle(\delta\rho_s)^2\rangle/2$ in our model as equal to
the quantity $g\Lambda^2/\pi$ in the model of Ref.\ \cite{han}.

Next, we consider the case where both $\rho_s$ and $\sigma_n$
have weak spatial variations.  We assume that $\sigma_n$
is given by eq.\ (12), that $\rho_n$ is spatially
varying, but not $\tau_n$, and that $\rho_n$ and $\rho_s$ are 
correlated according to the rule 
$\delta\rho_n  = \lambda\delta\rho_s$.
Then $\sigma_e$ is given
approximately by eq.\ (6), with 
\begin{equation}
\sigma_{av} = \frac{\rho_{n,av}\tau_n}{1 - i\omega\tau_n}
+ \frac{i\rho_{s,av}}{\omega} 
\label{eq:fluc2a}
\end{equation}
and
\begin{equation}
\langle (\delta\sigma)^2\rangle
= \langle (\delta\rho_s)^2\rangle \left(\frac{i}{\omega} 
+ \frac{\lambda\tau_n}{1-i\omega\tau_n}\right)^2.
\label{eq:fluc2b}
\end{equation}

Some representative plots of 
$\sigma_{e,1}(\omega)$ resulting from
eqs.\ (\ref{eq:weak}), (\ref{eq:fluc2a}) and (\ref{eq:fluc2b}) are 
shown in Fig.\ 1 for $\lambda = 1$ (perfect correlation) 
$\lambda = -1$ (perfect anticorrelation), and $\lambda = 0$
(no fluctuations in $\delta\rho_n$).  The case $\lambda = -1$
appears most plausible physically, since it implies that
the sum $\rho_n + \rho_s$ is non-fluctuating.  It also produces
the largest increase in $\sigma_{e,1}$ at any given $\omega$.

For $\omega\tau_n \ll 1$, eqs.\
(\ref{eq:weak}), (\ref{eq:fluc2a}), and (\ref{eq:fluc2b})
lead to the analytical result
\begin{equation}  
\sigma_{e,1}(\omega) \approx \sigma_{n,av} +
\frac{\langle(\delta\rho_s)^2\rangle}{2}
\frac{\sigma_{n,av} - 2\lambda\tau_n\rho_{s,av}}
{\rho_{s,av}^2 + \omega^2\sigma_{n,av}^2}.
\label{eq:25}
\end{equation}
Here $\sigma_{n,av} \equiv \rho_{n,av}\tau_n$ and we
have neglected corrections of order $(\omega\tau_n)^2$ in
the numerator.  Eq.\ (\ref{eq:25}) is again identical 
with Han's result\cite{han}, obtained using a replica formalism, 
if we make the identification 
$\langle(\delta\rho_s)^2\rangle/2 \leftrightarrow g\Lambda^2/\pi$.   

\subsection{Temperature-Dependent Superfluid Density; 
Effective-Medium Approximation}

Next, we calculate $\sigma_e(\omega)$ using a simple model for the 
conductivity of a CuO$_2$ layer proposed in Ref.\
{\it et al}\cite{corson2}, but generalized to include 
inhomogeneities.  We assume that $\sigma(\omega, {\bf x})$ is 
given by eq.\ (\ref{eq:model}), but for simplicity that it
can have only two values, $\sigma_A$ and $\sigma_B$, with 
probabilities $p$ and $1-p$.  
The corresponding conductivities are $\sigma_j(\omega)$ 
(j = A, B), where
\begin{equation}
\sigma_j(\omega) = \sigma_n(\omega) + i\frac{\rho_{s,j}}{\omega} 
\label{eq:condj}
\end{equation}
Following Ref.\ \cite{corson2}, we take 
$\sigma_n(\omega) = \rho_n\tau_n/(1-i\omega\tau_n)$.  
We assume $\rho_n = \alpha T$, for $T < T_c$, as expected
for nodal quasiparticles in a $d_{x^2-y^2}$ 
superconductor, and  
$\rho_n(T>T_c) = \alpha T_c$. 
We also take $\tau_n^{-1} = \beta T$.  This linear 
$T$-dependence may be expected if $1/\tau_n$ is 
determined primarily by quasiparticle-quasiparticle 
scattering in two dimensions\cite{qp}, rather than impurity
scattering, which might give a $T$-independent $1/\tau_n$.
For the superfluid density, we 
write $\rho_{s,A} = \gamma \rho_{s,0}$,
$\rho_{s,B} = \gamma^{-1} \rho_{s,0}$, where 
$\rho_{s,0}(T) = \rho_{s,0}(0)\sqrt{1-2\alpha T/\rho_{s,0}(0)}$.
This form ensures that $\rho_{s,0}$ 
decreases linearly with $T$ at low $T$, as observed 
experimentally\cite{linear}, and that it vanishes at
a critical temperature $T_c$ as $\sqrt{T_c - T}$.
We use the experimental values of the parameters $T_c$, 
$\sigma_0 \equiv \rho_n\tau_n$, 
$\rho_{s,0}(0)$, $\alpha$ and $\beta$ 
(as quoted in Ref.\cite{corson2}) for our numerical estimates.
Finally, we have arbitrarily assumed that 
the inhomogeneity parameter $\gamma = 3$.

We compute $\sigma_e(\omega)$ using the Bruggeman 
effective-medium approximation (EMA)\cite{bruggeman}:
\begin{equation}
p\frac{\sigma_A(\omega)-\sigma_e(\omega)}{\sigma_A(\omega)+\sigma_e(\omega)}
+(1-p)
\frac{\sigma_B(\omega)-\sigma_e(\omega)}{\sigma_B(\omega)+\sigma_e(\omega)} = 0.
\label{eq:ema1}
\end{equation}
The physically relevant solution is obtained
by requiring that $\sigma_e(\omega)$ be continuous in $\omega$ and 
that $\sigma_{e,1}(\omega) > 0$.
We also arbitrarily assume that $p=1/2$.
Our form for $\rho_{s,A}$ and $\rho_{s,B}$ guarantees
that the EMA solution 
$\rho_{s,e}(p = 1/2) =\sqrt{\rho_{s,A}\rho_{s,B}} = \rho_{s,0}$.

Fig.\ 2 
shows the resulting $\sigma_{e,1}(\omega, T)$ for several
frequencies ranging from 0.2 to 0.8 THz, the range measured in Ref.\
\cite{corson2}. 
Also shown are 
$\int_{\omega_{min}}^{\omega_{max}}\sigma_{e,1}(\omega,T)d\omega$ 
for $\omega_{min}/(2\pi) = 0.2$THz and $\omega_{max}/(2\pi) = 
0.8$THz.  Finally, we plot
$\sigma_{n,1}(\omega, T)$ for these frequencies,
as well as
$\int_{\omega_{min}}^{\omega_{max}}\sigma_{n,1}d\omega$.
Clearly, $\sigma_{e,1}(\omega)$ is substantially increased 
beyond the quasiparticle contribution, because of 
spatial fluctuations in $\rho_s$.

\subsection{Percolation Effects}

We next consider $\sigma_{1,e}(\omega)$ near a percolative
superconductor to normal (S-N) transition.  Such a transition might
occur, for example, in a single CuO$_2$ layer doped by a 
non-superconducting element such as Zn, as has already been
discussed by several authors\cite{chai}.  
We assume $\sigma({\bf x})$ to be given by eq.\ 
(\ref{eq:model}), with a position-independent Drude  $\sigma_n$ 
given by eq.\ (\ref{eq:Drude}),
and a $\rho_s({\bf x})$ which has one of two values, $\rho_s$ or
zero, with probabilities $p$ and $1 - p$.  We calculate
$\sigma_e(\omega)$ using the EMA, eq.\ (\ref{eq:ema1}).
The EMA predicts an S-N transition at $p = 0.5 = p_c$.
For all $p$ on either side of $p_c$, $\sigma_{e,1}(p, \omega)$ 
has a peak at $\omega = 0$, whose height diverges and whose
half-width decreases to zero, as $p \rightarrow p_c$.
This behavior
is illustrated in Fig.\ 3, where we plot $\sigma_{e,1}(p, \omega)$
for several $p > p_c$, as calculated in the EMA.  
The other parameters are indicated in the Figure
caption.  The behavior of $\sigma_{e,1}$ is qualitatively 
similar to that shown in Fig.\ 2, but there is 
additional critical behavior near the
S-N transition, which is absent from the 
weakly disordered system.  

More exact results near $p_c$ can be obtained with the
help of a standard scaling hypothesis\cite{bergman,stroud82}.  
For the present model, it takes the form
\begin{equation}
\frac{\sigma_e}{\sigma_>} = |\Delta p|^tF_{\pm}
\left(\frac{\sigma_</\sigma_>}{|\Delta p|^{s + t}} \right).
\end{equation}
Here $\sigma_<$ and $\sigma_>$ denote the complex 
conductivities in the
S and N regions of the layer;
$\Delta p = p - p_c$, $s$ and $t$ are standard percolation
critical exponents, which
depend on dimensionality, and possibly on 
other structural details of the layer; and 
$F_{\pm}(x)$ are characteristic scaling functions which
apply above and below $p_c$.  The expected behavior\cite{bergman}
is $F_+(x) \sim 1$ for $x \ll 1$, and $F_-(x) \sim x$  
for $x \ll 1$, while $F_{\pm}(x) \sim x^{t/(s+t)}$
for $x \gg 1$.  From these forms of the scaling functions,
we can infer\cite{stroud82} that $\sigma_{e,1}$ has a peak at $\omega = 0$
whose half-width on both sides of the percolation threshold
is
\begin{equation}
|\Delta\omega| \approx \frac{\rho_s}{\sigma_n}|\Delta p|^{s+t},
\end{equation}
which vanishes at the percolation threshold, where
$|\Delta p| \rightarrow 0$.  The height of this zero-frequency
peak diverges according to the law
$\sigma_{e,1}(p, \omega = 0) \rightarrow |\Delta p|^{-s}$
on either side of the percolation threshold.  For two-dimensional
bond percolation, $s = t \approx 1.30$, while in the EMA in any
dimension, $s = t = 1$.  Thus, we expect that 
$\sigma_{e,1}(p, \omega)$ is characterized near $p_c$ by a
line of diverging height, and half-width which goes
to zero near $p_c$.  Precisely this type of behavior is seen
in the EMA curves of Fig.\ 3.

\subsection{Tensor Conductivity}

The results shown in Fig.\ 3 correspond to an in-plane scalar 
superfluid density $\rho_s({\bf x})$ with a random spatial variation.
If instead $\rho_s({\bf x})$ were a $2 \times 2$
{\em tensor} with principal axes varying randomly with position
(as might occur for random quenched domains of stripes), a 
similarly enhanced $\sigma_{e,1}$ would be expected.  
If the two principal components of
the conductivity tensor are given by eq.\ (\ref{eq:condj}), 
and the stripes can point with equal probability
along either crystallographic direction, 
then $\sigma_e(\omega)$ is given in the EMA
by eq.\ (\ref{eq:ema1}) with $p = 0.5$.   
The resulting $\sigma_{e,1}(\omega)$ would behave exactly like that 
shown in Fig.\ 2.

\section{Discussion}

It is useful to compare our results to those of previous workers,
considering first the work of Han\cite{han}.  
For weak fluctuations in $\rho_s$, and a 
frequency-independent non-fluctuating $\sigma_n$, our
results are identical to his, provided we identify 
our parameter $\langle(\delta\rho_s)^2\rangle/2$ with his 
quantity $g\Lambda^2/\pi$.  However, our formal derivation
is more elementary than that of Ref.\ \cite{han}.
Because of this simplicity, we can easily 
consider a wide range of quenched inhomogeneities.
Since the two approaches
yield identical results for comparable types of 
inhomogeneities, we infer that our calculations
(which are based on standard methods of treating random
inhomogeneities in classical field equations such as those
of electrostatics) can also be done using field-theoretic
techniques.  This field-theoretic approach may therefore
be useful in treating other classical problems in 
heterogeneous media.

We briefly discuss relation between the present model and
that of Ref.\ \cite{barabash}.  In that earlier paper,
the CuO$_2$ layer was treated as an array of
small ``grains'' 
coupled together by overdamped resistively 
shunted Josephson junctions with Langevin thermal noise; 
$\sigma_e(\omega, T)$ was then computed using the classical 
Kubo formalism.  It consists of two parts: a sharp peak near the 
Kosterlitz-Thouless transition at $T_c$, due to breaking
of vortex-antivortex pairs, and a broader contribution, 
for $T < T_c$, which exists
only if the critical currents have quenched randomness.
This second contribution corresponds to that
considered here for spatially varying $\rho_s$.
For $T$ well below $T_c$, there are few vortices 
in the Josephson array; hence, the Josephson links
in the model of Ref. \cite{barabash} behave like inductors, and
the array acts like an inhomogeneous $LR$ network.  This  
network is simply a discretized version of the two-fluid 
model considered here: the resistances $R$ represent 
the normal quasiparticle channel,
and the inductances $L$ corresponds to the superfluid.

To summarize, we have shown that, in a 
superconducting layer with a spatially
varying $\rho_s$, there is extra absorption 
beyond that from low-lying quasiparticles alone.
Such quasiparticles are expected in a $d_{x^2-y^2}$ 
superconductor.  For several models of weak inhomogeneity, 
we have shown that the extra spectral weight at finite frequencies 
is proportional to the quantity 
$\langle(\delta\rho_s)^2\rangle/\rho_s$, 
as reported for $T < T_c$ in underdoped samples of 
BiSr$_2$Ca$_2$Cu$_2$O$_{8+x}$\cite{corson2}. 
Furthermore, in this regime, the extra weight 
generally appears as a Lorentzian centered at
zero frequency, with half-width proportional to $\rho_s/\sigma_n$.
Similar behavior is predicted for tensor inhomogeneities,
as is expected in stripe geometries.   We also predict that,
near an inhomogeneous superconductor-normal transition, 
the height of the inhomogeneity peak in $\sigma_1(\omega)$
diverges while its half-width goes to zero.  Thus, superfluid
inhomogeneity may be the principal source of the extra absorption
beyond the two-fluid model which is reported in the high-T$_c$
cuprates.

\section{Acknowledgments}

This work has been supported by NSF Grant
DMR01-04987, and by the U. S./Israel Binational Science Foundation.  
We are most grateful for useful conversations with Prof.\ David Bergman.

\newpage

\begin{figure}[hbt]
\epsfig{file=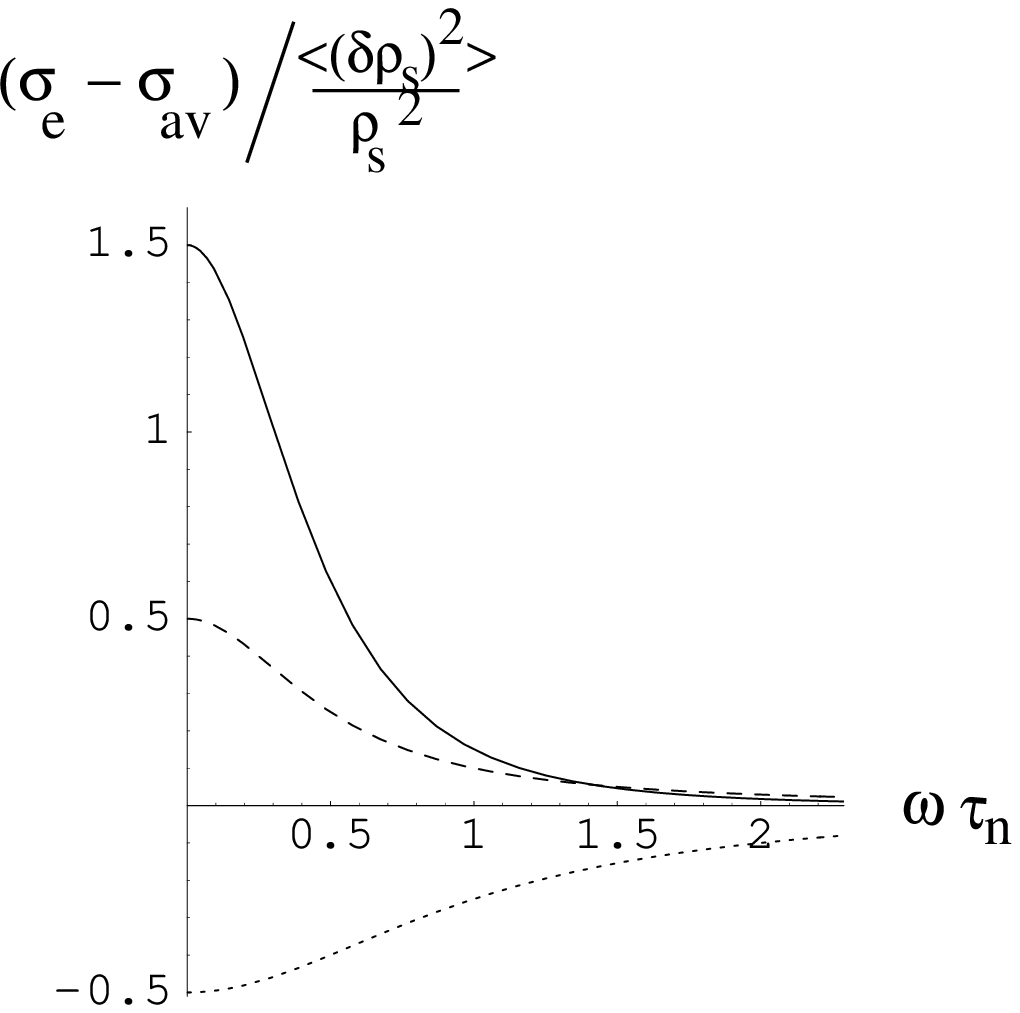, width=3in}
\vskip1pc
\caption{Excess real part of effective conductivity,
$\sigma_{e,1}(\omega) - \sigma_{n,1,av}(\omega)$, 
when there are weak correlated 
spatial fluctuations in both $\rho_s$ and $\rho_n$, normalized by the
relative magnitude of such fluctuations.  
$\sigma_{n,1,av} = \rho_{n,av}\tau_n/[1 + \omega^2\tau_n^2]$ is
the normal-state contribution.
We assume $\delta\rho_n = \lambda\delta\rho_s$, and show results
for $\lambda = -1$ (negative correlation, solid line),
$\lambda = 1$ (positive correlation, dotted line), 
and $\lambda = 0$ (no fluctuations in the normal fluid density, dashed line).  
The average superfluid and
normal fluid densities, $\rho_{s,av}$ and $\rho_{n,av}$, are 
assumed equal; the conductivities are in units of the quasiparticle
$\rho_n\tau_n$.}
\vskip5pc
\end{figure}

\begin{figure*}[hbt]
\epsfig{file=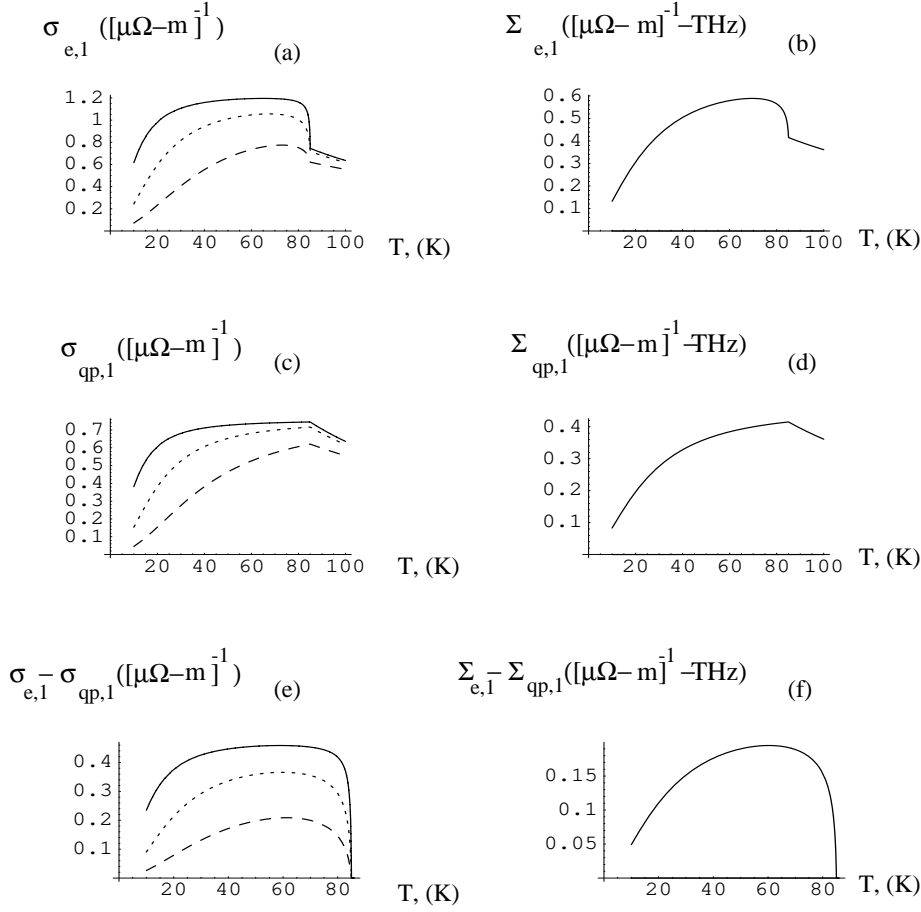, width=5in}
\vskip1pc
\caption{
(a) $\sigma_{e,1}(\omega, T)$, 
for $\omega/(2\pi) = 0.2$ THz (solid line), $0.4$ THz (dotted line), 
and $0.8$ THz (dashed line), 
for the model inhomogeneous superconductor described in the text.  
Also plotted are 
(b) $\Sigma_{e,1}\equiv
\int_{\omega_{min}}^{\omega_{max}}\sigma_{e,1}(\omega)d\omega$,
(c) $\sigma_{n,1}(\omega,T)$,
(d) $\Sigma_{n,1}\equiv
\int_{\omega_{min}}^{\omega_{max}}\sigma_{n,1}(\omega)d\omega$, 
(e) $\sigma_{e,1}(\omega, T) -\sigma_{n,1}(\omega,T)$, and 
(f) $\Sigma_{e,1} -\Sigma_{n,1}\equiv
\int_{\omega_{min}}^{\omega_{max}}
[\sigma_{e,1}(\omega)-\sigma_{n,1}(\omega)]d\omega$, 
where $\omega_{min}/(2\pi) = 0.2$Thz and 
$\omega_{max}/(2\pi) = 0.8$THz.
}
\vskip1pc
\end{figure*}

\begin{figure*}[t]
\epsfig{file=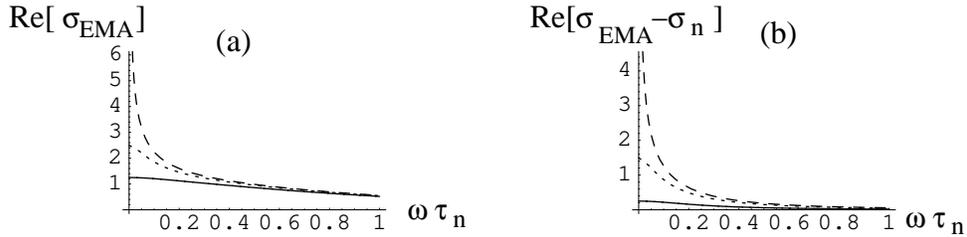, width=5in}
\vskip1pc
\caption{Real part $\sigma_{e,1}(p,\omega)$, in units of the quasiparticle
$\rho_n\tau_n$, of the effective
conductivity of a two-dimensional composite of normal metal
and superconductor, as calculated using the
two-dimensional effective-medium approximation for several
values of $p \geq p_c$: $p=0.9$ (solid line), $p=0.7$ (dotted line) and
$p=0.5$ (dashed line).}
\vskip1pc
\end{figure*}


\begin{thebibliography}{99}
\bibitem{remark} Eq.\ (\ref{eq:sigesn}) is actually valid for any 
$\sigma_n$ satisfying at large frequencies 
$\sigma_n(\omega) \sim i\rho_n/\omega$.
\end{thebibliography}
\end{document}